\documentclass[apj]{emulateapj}

\def\Journal#1#2#3#4{{#4}, {#1}, {#2}, #3}

\def\AAA{A\&A}
\def\ApJ{ApJ}
\def\AJ{Astronom. Journal}

\begin{document}

\title{Break in the VHE spectrum of PG 1553+113: new upper limit on its redshift?}

\author{Daniel Mazin\altaffilmark{a} \& Florian Goebel\altaffilmark{a}}
\altaffiltext{a} {Max-Planck-Institut f\"ur Physik, D-80805 M\"unchen, Germany; 
email: mazin@mppmu.mpg.de}

\begin{abstract}

PG\,1553+113 is a known BL Lac object, newly detected in the GeV-TeV energy
range by H.E.S.S and MAGIC. The redshift of this source is unknown and a lower
limit of $z > 0.09$ was recently estimated. The very high energy (VHE) spectrum
of PG\,1553+113 is attenuated due to the absorption by the low energy photon
field of the extragalactic background light (EBL). Here we correct the combined
H.E.S.S and MAGIC spectrum of PG\,1553+113 for this absorption assuming a
minimum density of the evolving EBL. We use an argument that the intrinsic
photon index cannot be harder than $\Gamma = 1.5$ and derive an upper limit on
the redshift of $z < 0.69$. Moreover, we find that a redshift above $z = 0.42$
implies a possible break of the intrinsic spectrum at about 200\,GeV.  Assuming
that such a break is absent, we derive a much stronger upper limit of $z <
0.42$. Alternatively, this break might be attributed to an additional emission
component in the jet of PG\,1553+113. This would be the first evidence for a
second component is detected in the VHE spectrum of a blazar.

\end{abstract}


\section{Introduction}

PG\,1553+113 was discovered in the Palomar-Green survey of UV-excess stellar
objects \citep{green}.  It is classified as a BL Lac object based on its
featureless spectrum \citep{redshift1, redshift2} and its significant optical
variability \citep{variability}.  PG\,1553+113 is well studied from the radio
to the X-ray regime.  Based on its broad-band spectral energy distribution
(SED), PG\,1553+113 is now classified as a high-frequency peaked BL Lac
\citep{classification}, similar to most of the AGNs detected at very high
energies (VHE).

The redshift of PG\,1553+113 is essentially unknown.  It was initially
determined to be $z$=$0.36$ \citep{redshift1} but later this claim was
withdrawn \citep{redshift2}.  To date no emission or absorption lines have been
measured despite several observation campaigns with optical instruments. No
host galaxy was resolved in {\it Hubble Space Telescope} (HST) images of
PG\,1553+113 taken during the HST survey of 110 BL Lac objects
\citep{Hubble_image}.  The HST results were used to set a lower limit of
$z$$>$0.78 \citep{z_extreme} for PG\,1553+113.  A more recent publication
claims a lower limit on the redshift of $z$$>$0.09 using the ESO-VLT optical
spectroscopy survey of 42 BL Lacertae objects of unknown redshift
\citep{z_new}.  The possibility of a large redshift is of critical importance
to VHE observations due to the absorption of VHE photons \citep{EBL_effect,
EBL_effect2} by pair-production on the extragalactic background light (EBL).
This absorption, which is energy dependent and increases strongly with
redshift, distorts the VHE energy spectra observed from distant objects. It is
believed that VHE photons above $E>100\,$GeV from objects at distances $z>1$
are completely absorbed. Note that the most distant  objects detected so far at
VHE are 1ES\,1218+304 ($z=0.182$, \citet{magic1218}) and 1ES\,1101-232
($z=0.186$, \citet{hessebl}).

Recently, VHE $\gamma$-ray emission from PG\,1553+113 was measured by H.E.S.S
\citep{hess1553} and by MAGIC \citep{magic1553}.  Both collaborations reported
a soft energy spectrum with a differential photon index of $\Gamma = 4.0 \pm
0.6$ and $\Gamma = 4.2 \pm 0.3$ respectively.  The VHE data from the two
measurements were used independently to derive an upper limit on the redshift
of the source of $z < 0.74$. This limit is based on the assumption of a minimum
possible level of the EBL and the maximum hardness of the intrinsic VHE
spectrum.  Although the intrinsic VHE $\gamma$-ray spectra of the AGNs are not
well known, it can be assumed that the intrinsic photon index of the sources is
not harder than $\Gamma_{\mathrm{int}}$=1.5 \citep{hessebl}.  However, there
are emission scenarios where the maximum possible VHE photon index is even
harder ($\Gamma_{\mathrm{int}}=2/3$, \citet{katar}).  

Here we use the combined H.E.S.S. and MAGIC spectrum of PG\,1553+113 to derive
upper limits on its redshift, assuming the two limits on the hardness of the
intrinsic photon index ($\Gamma_{\mathrm{int}}$=1.5 and
$\Gamma_{\mathrm{int}}=2/3$).

In addition, we present an alternative method to estimate an upper limit on the
redshift of PG\,1553+113 assuming that there is no break in the intrinsic VHE
spectrum of the source. 


\section{Minimum EBL}
\label{sec:ebl}

\begin{figure*}
\begin{center}
\includegraphics[width=0.91\textwidth,angle=0,clip]{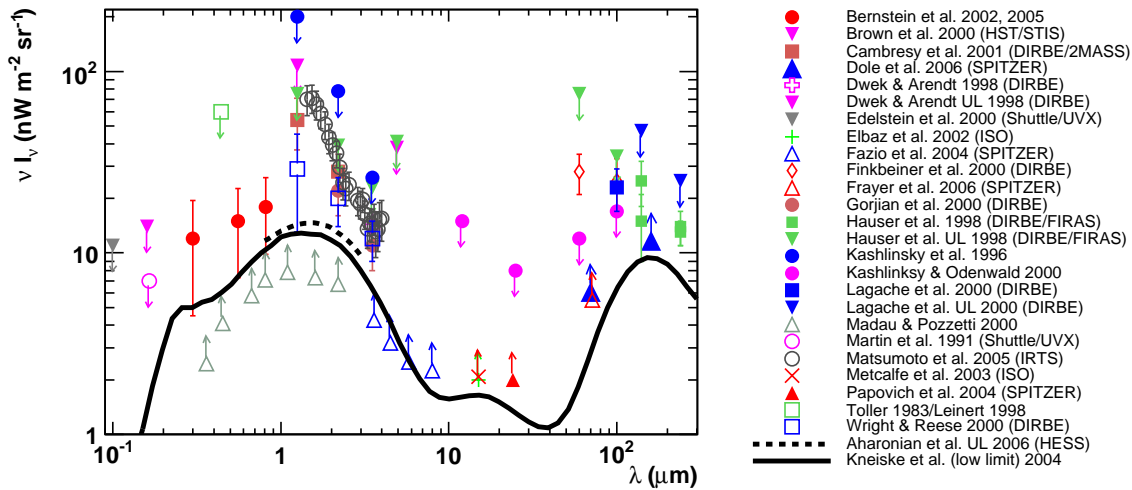}
\caption{Energy density of the extragalactic background light (EBL).
         Direct measurements, galaxy counts, low and upper limits are shown by
         different symbols as described in the legend. The dashed black line
         represents the upper limit set by H.E.S.S. \citep{hessebl}. The black
         solid curve is the minimum EBL spectrum taken from \citet{kneiske} for z=0.
         The model takes evolution of the EBL into account.}
\label{fig:EBLmin}
\end{center}
\end{figure*}

The most likely reaction channel in the interaction of VHE $\gamma$-rays with
the low energy photons of the EBL is pair production $
\gamma_{\mbox{\scriptsize{\,VHE}}} + \gamma_{\mbox{\scriptsize EBL}}
\rightarrow e^{+}\,e^{-}$. This reaction has its largest cross section when the
center of mass energy is roughly 2 times the threshold energy of 2m$_e$c$^2$.
The relevant EBL wavelength range for the absorption of VHE $\gamma$-rays spans
from UV light (0.1 $\mu$m) to far infrared (few 100 $\mu$m).  This light
predominantly  consists of redshifted star light of all epochs and reemission
of a part of this light by dust in galaxies.  The intrinsic (de-absorbed) VHE
photon spectrum, $dN/dE_i$, of a blazar located at redshift $z$ can be
determined using: \[ dN/dE_i\,=\, dN/dE_{obs} \times
\exp[\tau_{\gamma\gamma}(E,\,z)], \] where $dN/dE_{obs}$ is the observed
spectrum and $\tau_{\gamma\gamma}(E,\,z)$ is the optical depth.  While the
cross-section of the pair production is well known the spectral energy density
of the EBL is not, which makes the calculation of the optical depth uncertain.
Direct measurements of the EBL are difficult because of the strong foreground
emission, mainly consisting of reflected sunlight and thermal emission from
zodiacal dust particles.  Indirect upper limits on the EBL density were
obtained using fluctuation analyses of the detected background radiation
\citep{kashlinsky96,kashlinsky00}.  Recently, an upper limit on the EBL density
in the range from 0.8 to 3 $\mu$m was determined using measured VHE spectra of
distant AGNs \citep{hessebl}.  On the other hand, a robust lower limit on the
EBL is set by galaxy counts in different wavebands
\citep{madau,spitzer,elbaz,metcalfe,papovich}.  A summary of current direct and
indirect measurements of EBL is given in Fig.~\ref{fig:EBLmin}.

In the present study, we want to use the lowest possible realistic level of the
EBL in order to derive a minimum redshift, above which a break in the VHE
spectrum of PG\,1553+113 becomes evident.  Several EBL models have been
developed, taking into account the evolution of the EBL, which is related to
star and galaxy evolution (e.g. \citet{kneiske,primack,steckerEBL}).  Though
different in approach, the models agree in their predictions within a factor of
2.  For this study, we use the lower limit model from \citet{kneiske}, which is
just at the level of the direct lower limits set by the galaxy counts (see
Fig.~\ref{fig:EBLmin}).


\section{Combining the spectra of PG 1553+113 from H.E.S.S. and MAGIC}
\label{sec:comb}

\begin{figure}
\begin{center}
\includegraphics*[width=0.42\textwidth,angle=0,clip]{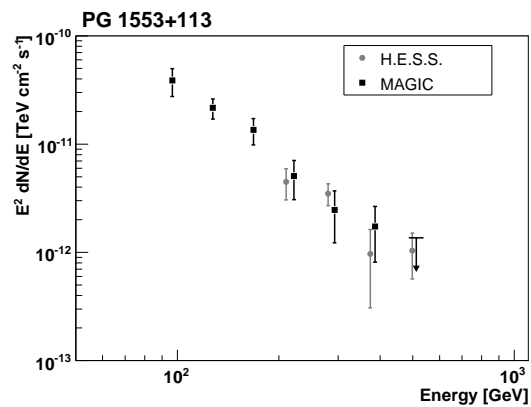}
\caption{Differential measured energy spectrum of PG 1553+113 
         multiplied by E$^2$ to represent energy density. 
         The spectrum measured by H.E.S.S. is shown by grey points, whereas
         the MAGIC spectrum is shown by black boxes. The last point in
         the MAGIC spectrum is a 95\% upper limit.}
\label{fig:spec}
\end{center}
\end{figure}

The differential VHE $\gamma$-ray spectra of PG\,1553+113 published by H.E.S.S.
and MAGIC are shown in Fig.~\ref{fig:spec}. The photon fluxes are multiplied by
$E^2$, which is equivalent to a $\nu F(\nu)$ representation. The H.E.S.S. data are shown
as grey circles and the MAGIC data by black boxes. 
For H.E.S.S., the data were taken in spring--summer 2005. In case of MAGIC, it
is a combined data set taken between spring 2005 and winter 2006. 
The last spectral point is a 95\% upper limit and will not be used in the further analysis.
AGNs are known to be variable sources in flux so that, in general, it is not trivial
to combine non-simultaneous data sets.  
In the present case, however, the agreement between
the spectra measured by H.E.S.S. and MAGIC is very good.  
As a test, we fitted a power law 
\begin{equation}
\label{eq:pl}
dN/dE = N_0 E^{-\Gamma} 
\end{equation} 
to the observed spectra in the
overlapping energy region between 150 and 600\,GeV by fixing the slope
(observed photon index $\Gamma$) to the value measured by H.E.S.S.: $\Gamma = 4$.
The resulting normalization factors are  $N_0 = (2.01 \pm 0.34) \times
10^{-13}\,\, \mathrm{phot} / (\mathrm{TeV}\,\mathrm{cm}^2\,\mathrm{s})^{-1}$
and $N_0= (2.24 \pm 0.59) \times 10^{-13}\,\, \mathrm{phot} /
(\mathrm{TeV}\,\mathrm{cm}^2\,\mathrm{s})^{-1}$ for the H.E.S.S. and MAGIC data
respectively. These numbers differ by 10\%, which is small compared to the
individual statistical errors.  In order to increase the
statistical power of the tests described below, we use the combined spectrum of
MAGIC and H.E.S.S. data on PG\,1553+113.
We normalize the spectra by multiplying the H.E.S.S. fluxes by a factor 1.1 to
avoid a possible bias in our results.  In order to show the effect of the
normalization we also perform the same study on a combined spectrum without
this normalization.


\section{Determination of upper limits on the redshift of PG\,1553+113}
\label{sec:stat}

Two different methods are used. The first one 
assumes that the spectral index of the intrinsic differential
photon spectrum is not higher than 1.5 
(or, alternatively, 2/3). The second one assumes that the intrinsic photon
spectrum has no break in the measured energy region from 80 to 600\,GeV.

\begin{figure}
\begin{center}
\includegraphics[width=0.42\textwidth,angle=0,clip]{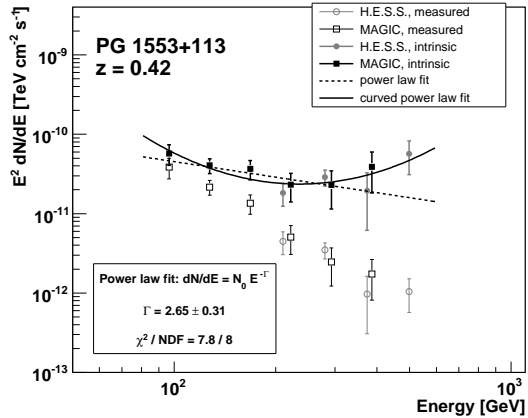}
\caption{Constraint on the redshift of PG 1553+113. 
{\it Hollow points}: measured combined (normalized) differential 
energy spectrum of PG\,1553+113 using MAGIC (squares) and H.E.S.S. (circles) 
data from 2005 and 2006. 
{\it Filled points}: intrinsic spectrum of PG\,1553+113, using
minimum possible density of the evolving EBL and the redshift of $z=0.42$.
{\it Black dotted line}: power law fit to the intrinsic spectrum; 
the fitted photon index is listed in the inlay.
{\it Black solid line}: curved power law fit.
The probability of the likelihood ratio test is 95.06\%, meaning that 
the fit by a curved power law is to be preferred over the fit by a power law.}
\label{fig:1553spec}
\end{center}
\end{figure}

\subsection*{Maximum intrinsic photon index $\Gamma_{\mathrm{int}}$}

A fit of  Eq.~\ref{eq:pl2} is performed to the 
intrinsic spectrum of PG\,1553+113. 
The fitted index $\Gamma_{\mathrm{int}}$ and its error $\sigma_{\Gamma}$ are combined to
$\Gamma_{\mathrm{max}}$, where $\Gamma_{\mathrm{max}} = \Gamma_{\mathrm{int}} + 2 \times \sigma_{\Gamma}$
corresponding to a 95\% confidence level. 
We note that despite the fact that for $z>0.42$ (see below) a curved power law gives a significantly better fit to the 
intrinsic spectrum than a simple power law, the $\chi^2$ value of
the latter fit is still acceptable.
A redshift $z$ is considered to be unrealistic if $\Gamma_{\mathrm{max}} < 1.5$ or,
in case of the extreme scenario, $\Gamma_{\mathrm{max}} < 2/3$.

\subsection*{Presence of a break in the intrinsic photon spectrum}

The second method is based on the indication that for larger redshifts 
($z>0.3$) the intrinsic spectrum of PG\,1553+113 exhibits a break at about
200\,GeV with the intrinsic energy spectrum rising after the break 
(see Fig.~\ref{fig:1553spec}).
Such a break in the intrinsic spectrum of PG\,1553+113
cannot be excluded {\it {a priori}}. 
It could attributed to a second population of $\gamma$-ray emitting particles.
However, in none of the
measured VHE $\gamma$-ray spectra of extragalactic sources such a component has
been found.  Thus, if a spectrum shows a break, either it is the first time a
second emitting component is found in a VHE $\gamma$-ray spectrum or a lower
redshift value $z$ has to be assumed.

To test the presence of a second component 
(or the presence of a break) 
in the VHE spectrum of
PG\,1553+113 we performed a likelihood ratio test \citep{stat2} on the intrinsic spectrum.  
Two hypotheses are tested.
Hypothesis A is a simple power law (2 free parameters):
\begin{equation}
\label{eq:pl2}
  dN/dE = N_0 E^{-\Gamma_{\mathrm{int}}}
\end{equation}
Hypothesis B is a curved power law (3 free parameters):
\begin{equation}
\label{eq:cpl}
  dN/dE = N_0 E^{-(\alpha\,+\,\beta\,\ln(E))} \,,
\end{equation}
which corresponds to a parabolic law in a 
$\log (\mathrm{E}^{2}\, \mathrm{dN}/\mathrm{dE})$ 
vs. $\log \mathrm{E}$  representation. A parabolic shape is a natural 
choice to describe the transition region of a break between two 
spectral components.
By fitting the two functional forms (Eq.~\ref{eq:pl2} and \ref{eq:cpl})
to the de-absorbed spectrum one obtains values
of the likelihood functions $L_A$ and $L_B$. If hypothesis A is true the likelihood
ratio $R = -\ln(L_A/L_B)$ is approximately $\chi^2$ distributed with one degree of freedom.
One defines a probability 
\begin{equation}
\label{eq:probab}
   P = \int\limits_{0}^{R_{\mathrm{meas}}} p\,(\chi^2) \, \mathrm{d\chi^2}
\end{equation}
where $p\,(\chi^2)$ is the $\chi^2$ probability density function and $R_{meas}$ 
is the measured value of $R$. Hypothesis A will be rejected 
(and hypothesis B will be accepted) if $P$ is greater than 
the confidence level, which is set to 95\%.


\section{Results}
\label{sec:res}

\begin{figure}
\begin{center}
\includegraphics[width=0.45\textwidth,angle=0,clip]{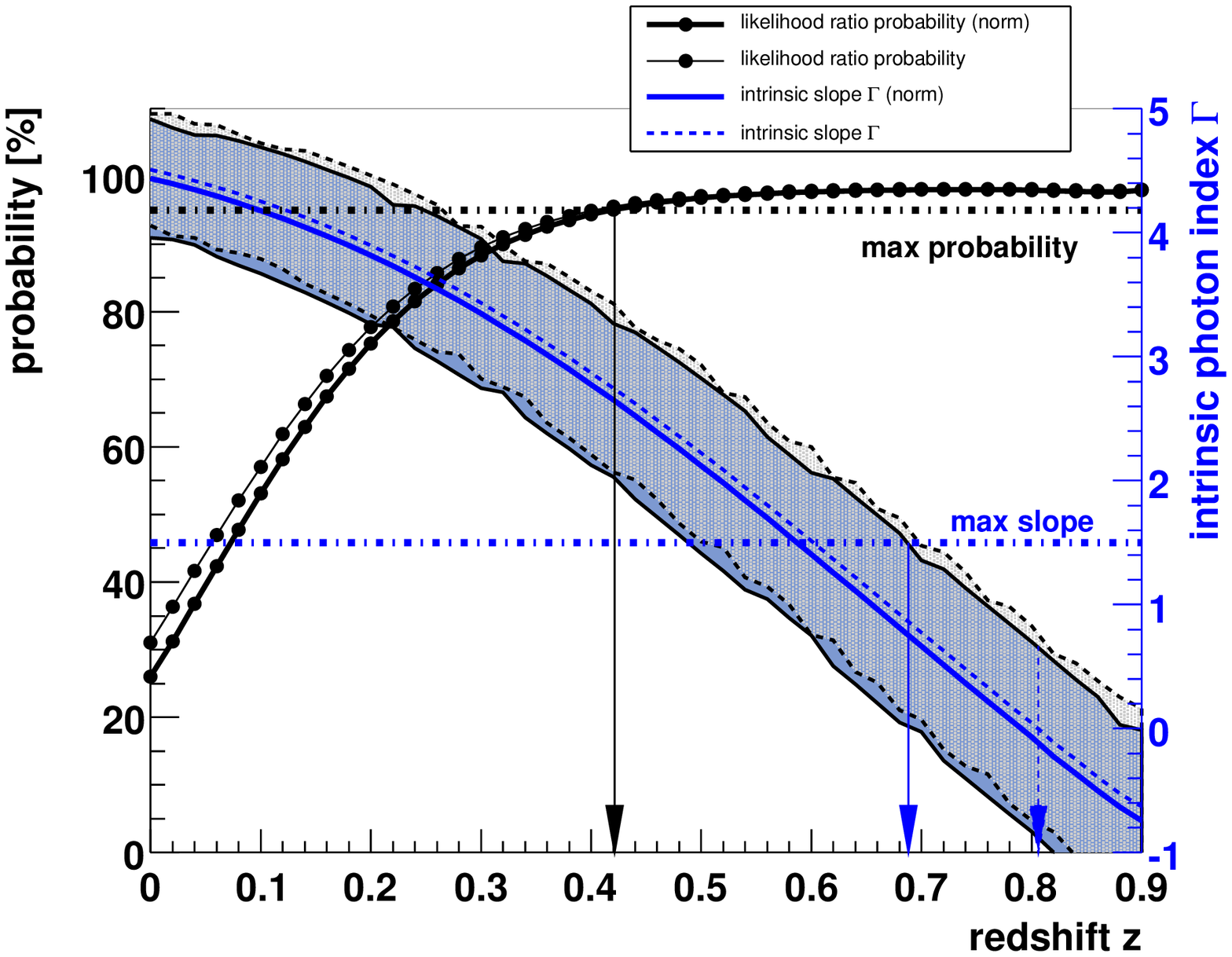}
\caption{Constraints on the redshift of PG 1553+113 (see text for details). \newline
{\it Black points} and left Y-axis: probability of the likelihood ratio test 
(Eq.~\ref{eq:probab}).
The black arrow indicates the minimum redshift ($z=0.42$), at which 
the break in the intrinsic spectrum of PG\,1553+113 is evident. \newline
{\it Blue line} and right Y-axis: intrinsic
photon index $\Gamma_{\mathrm{int}}$. 
Shaded areas correspond to 2 $\sigma$ confidence belt.
The intrinsic spectrum 
is harder than this for $z>0.69$, which is indicated by the blue arrow.
The extreme case of $\Gamma=2/3$ leads to an upper limit of
$z<0.80$ as indicated by the dashed blue arrow.}
\label{fig:limits}
\end{center}
\end{figure}

We examined a wide range of redshifts values $z$ between 0.1 and 0.9 in steps of
0.02. Each time, the corresponding optical depth was calculated and the intrinsic
spectrum of PG\,1553+113 was determined using the low limit model from \citet{kneiske}. 
The probability of the likelihood ratio 
test and the intrinsic photon index $\Gamma_{\mathrm{int}}$ as a function of 
redshift $z$ are shown in Fig.~\ref{fig:limits}.

The intrinsic photon index $\Gamma_{\mathrm{int}}$ as a function of the
redshift is shown by the thick blue line in Fig.~\ref{fig:limits}.
A $2\sigma$ confidence belt is drawn as blue shaded area. 
The result without the normalization between the H.E.S.S.
and MAGIC measured spectrum of PG\,1553+113 is shown by the dashed blue line with a 
corresponding $2\sigma$ confidence belt as grey shaded area.
Assuming that the intrinsic photon index can not be harder than $\Gamma_{\mathrm{int}}=1.5$, 
we obtain a redshift limit of $z<0.69$ with a confidence of 95\%. 
Assuming the maximally hard spectrum as proposed by \citet{katar} 
with $\Gamma_{\mathrm{int}}=2/3$, we obtain a redshift upper limit
of  $z<0.80$. 

The intrinsic spectrum was considered to have a break 
if the likelihood ratio test gave more 
than 95\% confidence.  The probability of the likelihood ratio test as a function
of the redshift is shown by the black points and thick black line in Fig.~\ref{fig:limits}. 
The smallest redshift, for which the test gave more than 95\% confidence, 
is $z=0.42$ (see Fig.~\ref{fig:1553spec}). The assumption that there is no 
break in the intrinsic spectrum of PG\,1553+113 thus leads to an upper limit on 
its redshift of $z<0.42$.   
The result without the normalization between the H.E.S.S.
and MAGIC measured spectrum is shown by the thin black line.
The systematic uncertainty in the derived limits, arising from the applied 
normalization of 10\% between the H.E.S.S. and MAGIC data, is below 10\%.

The test for the existence of a break cannot be applied to the  H.E.S.S. spectrum
alone since the break occurs around the lower end of the energy region of the H.E.S.S. measurement.
On the other hand, applying this test to the MAGIC data alone never reaches
the required probability of 95\% for the existence of the break due to insufficient
statistics.


\section{Conclusion and Outlook}
\label{sec:conc}

In this study, we combined H.E.S.S. and MAGIC data using their good agreement and used
a realistic minimum density of the EBL to reconstruct the intrinsic spectrum
of PG\,1553+113.
We showed that the intrinsic photon index $\Gamma_{\mathrm{int}}$ becomes smaller than
$\Gamma_{\mathrm{int}}=1.5$ at $z=0.69$, which can be considered a robust upper limit on the
redshift of PG\,1553+113. In case of the extreme emission scenario with $\Gamma_{\mathrm{int}} = 2/3$,
we obtain an upper limit of $z<0.80$.
Moreover, we showed that a break in the intrinsic spectrum at about 200\,GeV becomes evident
at a redshift of $z=0.42$. The break can either be interpreted
as an upper limit on the redshift of PG\,1553+113 or as evidence for a second emission 
component in the VHE spectrum of the object. 
The upper limit of $z<0.42$ implies values of the intrinsic slope indicating that the peak 
of the high-energy component of the SED lies below few hundred
GeV, as typically derived for the closest TeV blazars Mrk\,421 and Mrk\,501 (in low flux state).
We note that increasing the statistics by combining the spectra of MAGIC and 
H.E.S.S. resulted in a moderate improvement of the redshift upper limit.
On the other hand, the second method, which is based on 
the search for structures in the intrinsic spectrum and which resulted in much more
stringent limits, became only feasible using the combined data set.

The knowledge of the distance of PG\,1553+113 is crucial for the modeling
of the emission processes in the object, especially for combining X-ray and VHE data. 
In case the redshift of PG\,1553+113 is larger than $z>0.2$ it would not only be
the farthest AGN detected in the VHE range so far but also by far the most energetic one.
Further campaigns to determine the redshift of PG\,1553+113 directly as well as extensive
VHE observations are encouraged.

\vspace{-2mm}

\acknowledgements
{ 
We would like to thank Wolfgang {\mbox{Wittek}} for fruitful 
discussions and help in preparing this manuscript, as well as 
Wystan Benbow for providing H.E.S.S. data of PG\,1553+113.
The financial support of  MPI f\"ur Physik, Munich, is acknowledged.}

\end{document}